# Multiplicative and additive modulation of neuronal tuning with population activity affects encoded information

Iñigo Arandia-Romero[1,2], Seiji Tanabe[3], Jan Drugowitsch[4], Adam Kohn[3] and Rubén Moreno-Bote[1,2,5,6]

[1] Department of Information and Communication Technologies, Universidad Pompeu Fabra, 08018, Barcelona, Spain
[2] Research Unit, Parc Sanitari Sant Joan de Deu, Esplugues de Llobregat, Barcelona, Spain, 08950
[3] Dominick Purpura Department of Neuroscience and Ophthalmology and Visual Science, Albert Einstein College of Medicine, Bronx, New York, 10461
[4] Département des Neurosciences Fondamentales, Université de Genève, CH-1211 Geneva 4, Switzerland
[5] Centro de Investigación Biomédica en Red de Salud Mental (CIBERSAM), Esplugues de Llobregat, Barcelona, Spain, 08950
[6] Serra Húnter Fellow Programme, Universidad Pompeu Fabra, Barcelona, 08018, Spain

Corresponding author:
Rubén Moreno-Bote

Department of Information and Communication Technologies, Universidad Pompeu Fabra, 08018, Barcelona, Spain

Email: ruben.moreno@upf.edu

Acknowledgements: I.A.R is supported by a PhD. grant from the Department of Education, Linguistic Politics, and Culture of the Basque Government. AK is supported by the National Institutes of Health (EY016774), an Irma T. Hirschl Career Scientist Award, and Research to Prevent Blindness. R.M.B. is supported by the Ramón y Cajal Spanish Award RYC-2010-05952, the Marie Curie FP7-PEOPLE-2010-IRG grant PIRG08-GA-2010-276795, and the Spanish PSI2013-44811-P grant.

## Abstract

Numerous studies have shown that neuronal responses are modulated by stimulus properties, and also by the state of the local network. However, little is known about how activity fluctuations of neuronal populations modulate the sensory tuning of cells and affect their encoded information. We found that fluctuations in ongoing and stimulus-evoked population activity in primate visual cortex modulate the tuning of neurons in a multiplicative and additive manner. While distributed on a continuum, neurons with stronger multiplicative effects tended to have less additive modulation, and vice versa. The information encoded by multiplicatively-modulated neurons increased with greater population activity, while that of additively-modulated neurons decreased. These effects offset each other, so that population activity had little effect on total information. Our results thus suggest that intrinsic activity fluctuations may act as a 'traffic light' that determines which subset of neurons are most informative.

# Introduction

Neuronal activity fluctuates at both the single neuron and the population levels. These activity fluctuations can limit the reliability of neuronal codes because a given response can arise from several distinct sensory stimuli (Shadlen and Newsome, 1998; Tolhurst et al., 1983). Fluctuations in stimulus-evoked responses have been generally viewed as harmful noise that needs to be averaged out to extract the desired signal (Cohen and Maunsell, 2009; Mitchell et al., 2009; Shadlen and Newsome, 1998). Recent work has shown, however, that population activity fluctuations modulate single-cell stimulus-evoked responses in additive and multiplicative manners (Ecker et al., 2014; Goris et al., 2014; Lin et al., 2015; Scholvinck et al., 2015), suggesting that they are highly structured and hence might have a computational role. However, the role, if any, of fluctuations of total activity in neuronal populations on sensory neuronal tuning and encoding has not been demonstrated.

      We studied the influence of population activity fluctuations on the responses of single neurons and small neuronal ensembles in primary visual cortex (V1) of both anesthetized and awake monkeys. We found that the tuning for stimulus orientation of orientation-selective neurons changes multiplicatively or additively with the total, stimulus-evoked activity of the neuronal population that embeds these individual neurons, while leaving their tuning width and orientation preference mostly unaffected. While distributed on a continuum, neurons with strong multiplicative effects tended to have weak additive effects and vice versa, suggesting some specificity of the modulation across neurons. Consistent with a multi-gain model of neuronal responses, we found that neurons and small neuronal ensembles with strong multiplicative effects became more informative with stronger population activity, whereas those with strong additive effects became less informative. Population activity before stimulus onset was also predictive of both tuning modulation and changes in encoded information, but to a lesser degree than stimulus-evoked population activity. Importantly, we found that population activity does not substantially alter total sensory information in the recorded population. Rather, it routes how this information is represented, in an antagonist way, into multiplicatively and additively modulated neurons and neuronal ensembles. These results suggest that intrinsic fluctuations in the activity of neuronal populations may act as a 'traffic light'

that modulates the tuning of individual neurons and can differentially redistribute sensory information in the neuronal population.

## Results

We recorded neuronal populations in the superficial layers of V1 in four anesthetized (datasets 1-4, D1-D4) and one awake (D5) macaque monkeys. We measured responses to gratings drifting in 8 (12 for D5) equally spaced directions. Gratings were presented for 1,280 ms (350ms) each, interleaved with a 1,500 ms (50ms) blank screen and repeated 300 or 400 (50) times in random order. We analyzed the activity of 567 single neurons and multiunits, which we refer to together as 'units'. We analyzed 122, 106, 73, 161, and 18 simultaneously-recorded units in datasets D1 to D5, respectively. We also analyzed separately a subset of 83 well-isolated single neurons (27, 14, 7, 31 and 4 from D1-D5, of which 12, 12, 4, 15, 2 were orientation selective; see Experimental Procedures).

The firing rate of many V1 neurons is tuned to the orientation of a drifting grating (illustrated in Fig. 1A). Since neurons are embedded in a local network and are correlated (median of pair-wise spike count correlations: $\rho = 0.21$ across all anesthetized datasets), the summed total activity of that local population (called *population activity*) might modulate this tuning (Fig. 1B). This modulation could involve multiplicative or additive effects, or both, as well as broadening and displacement (Fig. 1C). Similarly, positive correlations among neurons can arise in multiple ways, such as additive modulation, multiplicative modulation, broadening of tuning, or a combination of these effects or others (Figure S1A). Therefore, the existence of correlations does not specify how tuning is modulated. We thus developed an analysis that could distinguish how tuning is modulated with population activity fluctuations.

In our data, population activity showed substantial fluctuations across trials for a fixed stimulus condition (several representative trials shown in Fig. 2A, left), consistent with previous reports (Arieli et al., 1996; Ecker et al., 2014; Scholvinck et al., 2015). The distribution of spike counts during the stimulation period across trials for one stimulus was roughly unimodal and broad (Fig. 2A, right; similar unimodal distributions were obtained in all datasets, Figure S2A). We characterized the timescale of the fluctuations using the spontaneous activity periods. Fluctuations in population activity were correlated with a timescale of a few hundreds of milliseconds (Figure S2B), consistent with previous reports on single neuron activity in V1 (Ecker et al., 2014; Kohn and Smith, 2005). Population activity was negatively correlated with LFP signals (Figure S2C), as previously reported (Okun et al., 2015).

We tested how neuronal tuning varies with fluctuations in population activity using a model-free approach, by comparing responses of a single neuron when the activity of the rest of the recorded neurons was either high (defined as the half of trials in which the summed population activity was greatest) or low (remaining trials). We used all recorded units to define periods of high and low population activity, excluding the neuron whose tuning was being characterized to avoid artifacts. Therefore, any observed modulation of tuning must arise from network effects, and would not be observed for uncorrelated neural populations. Both tuning and population activity were measured during the entire duration of the evoked activity period (shorter periods are considered below).

The tuning of an example single neuron depended clearly on population activity (Fig. 2B): responses were stronger when population activity was high (dark red box, Fig 2B) compared to when it was low (light red). To characterize how tuning was altered, we first determined whether there was substantial broadening (where tuning width was defined as the distance between peak to half-peak) or displacement of tuning with population activity. To quantify these effects, we fit the tuning of each neuron with a von Mises function (see Experimental Procedures). Across single neurons, we found a small (2% relative change) widening of tuning when population activity was high compared to low, but this effect was not significant (Figure S3A; Mann-Whitney $U = 855$, $p = 0.2$). Tuning preference was also only weakly modulated with population activity (Figure S3B; median absolute displacement = 1.0 degrees; permutation test $p < 0.002$), a small shift when compared to the typical tuning width. Therefore, we conclude that changes in tuning width and preference are small, and that the influence of population activity can only involve multiplicative and additive modulation of tuning.

**Multiplicative and additive modulation of tuning with population activity**

We sought to determine the extent to which tuning was multiplicatively and additively modulated with population activity. In the following analysis (Fig. 3A-C), both tuning and evoked population activity were measured from 160 to 260 ms after stimulus onset. This brief time period was chosen such that we could, on one hand, analyze the data from awake and anesthetized animals in the same way, and, on the other hand, study the temporal dynamics of the modulatory effects of population activity. The results for other time periods are discussed further below.

Tuning varied strongly with population activity (Fig. 3A, four examples shown). For each single neuron, we characterized its multiplicative and additive modulation with population activity by performing linear regression on the average response to each orientation, when population activity was high compared to when it was low (Fig. 3B). The slope of the linear fit indicates how tuning scales multiplicatively with population activity (termed hereafter the *multiplicative factor* (MF)). The intercept of the fit, on the other hand, describes the additive shift to tuning with population activity. To obtain a relative measure of the additive shift, like the multiplicative factor, we defined the *additive factor* (AF) as the ratio between this intercept and the mean firing rate of the neuron across orientations. Thus, neurons with purely multiplicative factors will feature a fit with slope larger than one that passes through the origin, whereas neurons with purely additive factors will have fits with slope one and a positive intercept. In a separate analysis, we confirmed that estimates of the multiplicative and additive factors from von Mises fits to the tuning gave similar results (not shown).

For the example single neuron of Fig 2B, tuning was modulated multiplicatively (MF = 1.4, permutation test $p < 0.002$) with little additive modulation (AF = 0.009, $p = 0.01$). The four example neurons of Fig 3A displayed different levels of multiplicative and additive modulation. For instance, the neuron in the 2$^{nd}$ panel was modulated in a mostly multiplicative manner (MF = 1.5, permutation test $p < 0.002$; AF = 0.046, $p = 0.04$), and the remaining neurons displayed a combination of multiplicative and additive effects. When we calculated tuning in more finely binned sets of trials, we observed that the modulation varied smoothly with the population activity level (Figure S3C).

Statistically-significant multiplicative and additive factors were found in a substantial fraction (26/45 for multiplication, 15/45 for addition) of orientation-selective single neurons. The median multiplicative factor across all single neurons was 1.28, significantly larger than one (Fig. 3C left, black; Mann-Whitney U = 215, $p = 10^{-10}$). This corresponds to a change of 28% in the firing rate, which occurs with a 35% increase in population activity between low-activity and high-activity trials. The median additive factor was also significantly larger than zero (Fig. 3C right, black; median = 0.06, Mann-Whitney U = 387, $p = 10^{-8}$), indicating a 6% increase relative to the neuron's mean firing rate at low population activity. Importantly, we found that there was a negative correlation between the multiplicative and additive factors across single neurons (Fig. 3D, black dots; $\rho = -0.48$, non-parametric bootstrap $p < 0.002$, see Experimental Procedures).

The results described thus far were based on well-isolated single neurons, but remained qualitatively unchanged if we included activity from multiunits. Across all units, the median multiplicative and additive factors were significantly larger than one and zero, respectively (Fig. 3C, white; median MF = 1.21, Mann-Whitney U = 2 $10^4$, $p = 10^{-48}$; median AF = 0.13, Mann-Whitney U = 2 $10^5$, $p = 10^{-44}$). The negative correlation between multiplicative and additive factors was also apparent across this large set of units (Fig. 3D, white dots: $\rho = -0.46$, $p < 0.002$). Although multiplicative and additive factors formed a continuum rather than distinct groupings, the negative correlation between multiplicative and additive factors across both single neurons and all units indicates a partial division of multiplicative and additive modulation with population activity.

To test whether the finding of both multiplicative and additive modulation with virtually no broadening was not due to artifacts in our estimation method, we applied the same method to simulated population activity with tuning identical to the one observed in the data (Figure S1). We created neuronal populations with purely multiplicative modulations, purely additive modulations, or purely broadening effects and tested whether our method discovered the true modulation while rejecting other types of modulation. The method reliably estimated the correct type of modulation in each simulated dataset (Figure S1B). We furthermore confirmed that the negative correlation between multiplicative and additive factors found in our data was not an artifact of our estimation method, as our method could reliably detect or reject the presence of correlations between these factors in simulated data (Figure S1C).

We also evaluated the significance of modulatory factors in each dataset separately, in part to test whether there are substantial differences between anesthetized and awake preparations. We found strong and significant multiplicative and additive factors in most individual datasets (Fig. 3E). The median multiplicative factors were significant in four of five datasets, including the awake dataset (Fig. 3E right; median = 1.28, permutation test $p < 0.002$, D1; median = 1.10, $p < 0.002$, D2; median = 1.02, $p = 0.4$, D3; median = 1.48, $p < 0.002$, D4; median = 1.36, $p = 0.04$, D5). Significant positive additive factors were found in all datasets (Fig. 3E right; median = 0.12, permutation test $p < 0.002$, D1; median = 0.2, permutation test $p < 0.002$, D2; median = 0.18, $p < 0.002$, D3; median = 0.07, $p < 0.002$, D4), except the awake dataset, for which there was a non-significant negative trend, presumably due to the lower number of neurons and trials recorded when compared to the anesthetized datasets (Fig. 3E right; median = -0.18, $p = 0.4$, D5). We also confirmed that the negative correlation between multiplicative and additive factors was present in all anesthetized datasets separately (Figure S3D), indicating that this correlation did not emerge from aggregating data with different mean values.

Substantial multiplicative and additive effects with no broadening, and a negative correlation between multiplicative and additive factors, were also observed when, instead of using direct measures of population activity, we used the projection of the population activity vector onto the first PCA component on a trial-by-trial basis (Figure S4A,B). Thus, our findings are not sensitive to the specific definition of population activity used but generalize to other sensible alternative measures of population activity strength.

Finally, we tested whether the tuning modulation was also present during other response epochs than the window 160-260 ms after stimulus onset, considered above. We repeated our analyses measuring both neuron tuning and population activity in 100 ms windows spanning the range from 60 to 1260ms. We found that the modulation of neuronal tuning with population activity was robust in these other epochs as well (Fig. 3F).

**Modulation of tuning with pre-stimulus population activity**

We have thus far considered how tuning changes with fluctuations in evoked population activity. These population fluctuations vary slowly under spontaneous conditions, over a timescale of hundreds of milliseconds (Figure S2B), and are well-documented (Arieli et al., 1996; Ecker et al., 2014; Fiser et al., 2004; Kenet et al., 2003; Kohn and Smith, 2005; Smith and Kohn, 2008; Tsodyks et al., 1999). Spontaneous activity fluctuations have been shown to influence subsequent evoked responses (Arieli et al., 1996; Tsodyks et al., 1999). Thus, we sought to determine how tuning during stimulus presentation varies with the strength of population activity before stimulus onset. We measured population activity in the 100 ms preceding stimulus onset, and tuning from 60 to 160 ms after stimulus onset. We excluded the data of the awake preparation, as the short inter-stimulus interval (50 ms) made a reliable estimation of pre-stimulus activity impossible.

Orientation tuning depended on the strength of pre-stimulus population activity. We found significant positive multiplicative and additive factors (Fig. 4A; for single neurons, median MF = 1.03, Mann-Whitney U = 645, $p$ = 0.01; median AF = 0.093, U = 430, $p$ = 5 $10^{-6}$; for all units, median MF = 1.06, U = 3 $10^4$, $p$ = 2 $10^{-16}$; median AF = 0.084, U = 2 $10^5$, $p$ = 3 $10^{-42}$). The modulation with pre-stimulus population activity was significantly smaller than that based on fluctuations in stimulus-evoked population activity (MFs: Wilcoxon sign-rank test, p < 0.01; AFs: p < 0.01). Furthermore, the modulation with pre-stimulus activity was most evident when tuning was measured shortly after stimulus onset (60-160ms; Fig. 4B). The factors typically declined over time, as one would expect from the spike correlation times of a few hundreds of milliseconds found in our data and usually reported for V1 (Arieli et al., 1996; Ecker et al., 2014; Fiser et al., 2004; Kenet et al., 2003; Kohn and Smith, 2005; Smith and Kohn, 2008; Tsodyks et al., 1999).

**A multi-gain model predicts how tuning modulation affects information encoding**

Our analysis revealed that tuning undergoes both multiplicative and additive modulation as a function of population activity, and that across neurons there is a negative correlation between these two types of modulation. To what extent does this tuning modulation influence encoded sensory information? Does information depend on whether the modulation was multiplicative or additive? To address these questions, we considered an idealized model with both multiplicative and additive tuning modulations. We assumed that the mean response of a neuron in the population depends on a global modulatory factor $g$ as

$$f_i(\theta, g) = \underbrace{g_{m,i}(g) h_i(\theta)}_{\text{multiplication}} + \underbrace{g_{a,i}(g)}_{\text{addition}} = \underbrace{(1+\alpha_i g) h_i(\theta)}_{\text{multiplication}} + \underbrace{\beta_i g}_{\text{addition}}. \tag{1}$$

The first term in the sums corresponds to the multiplicative modulation of tuning, and the second term corresponds to its additive modulation. The normalized tuning function $h_i(\theta)$ describes the tuning of the neuron with respect to the sensory variable $\theta$, which is modulated by the neuron-specific multiplicative and additive factors, $g_{m,i}$ and $g_{a,i}$, respectively. These factors relate to a global modulatory factor, $g$, linearly by $g_{m,i}(g) = 1 + \alpha_i g$ and $g_{a,i}(g) = \beta_i g$. For instance, a neuron with a purely multiplicative factor corresponds to $\alpha_i > 0$ and $\beta_i = 0$. The global modulatory factor $g$, assumed to be shared by all neurons in the population, generates correlations between neurons. Firing of each neuron, conditioned on the global modulatory factor, is assumed to be Poisson with the rate dictated by Eq. (1). This *multi-gain* model, with arbitrary mixtures of multiplicative and additive factors across neurons, is a generalization of recently introduced models with purely multiplicative modulation of neuronal variance and pair-wise covariance (Goris et al., 2014), or with purely additive modulation to describe state-transitions in neuronal populations (Ecker et al., 2014). Our model generalizes also the recent affine model (Lin et al., 2015), which allows arbitrary additive factors for each neuron but features a multiplicative factor identical for all neurons. In our model, in contrast, each neuron can have a different multiplicative factor $\alpha_i$ (see Equation 1), as our data suggest (Fig. 3). Using this more complex model was justified by its ability to better predict neural activity of a hold-out set than alternative models (Figure S5). Most of the models described above have not been used to make predictions about information encoding, and the predictions that have been made were not tested experimentally. The multi-gain model provides specific predictions about how sensory information in neural data should depend on the multiplicative and additive modulation of tuning, which we tested.

From our multi-gain model described in Eq. (1) it is straightforward to compute its Fisher information, which is a measure of discriminability between two nearby stimulus orientations (Ma et al., 2006; Seung and Sompolinsky, 1993). Because we are interested in how neurons' information about orientation depends on population activity, we conditioned information on the global modulatory factor $g$, resulting in

$$I_i(\theta, g) = \frac{g_{m,i}^2(g)\, h_i'^2(\theta)}{g_{m,i}(g)\, h_i(\theta) + g_{a,i}(g)}, \qquad (2)$$

where the prime denotes a derivative with respect to the stimulus (i.e. $h'(\theta)$ is proportional to the tuning slope). This equation captures the information provided by each neuron if there is no change in the relationship between response magnitude and variability. Consistent with this assumption, we found little difference in Fano factors between trials with low or high population activity (Figure S6). We also found similar correlations for the two sets of trials (Figure S6).

Eq. (2) predicts that a neuron's information about stimulus orientation increases with multiplicative gain (as their effect is dominated by the numerator), but decreases with additive modulation (as they only appear in the denominator). Intuitively, a multiplicative modulation increases the tuning slope, and thus information grows; in contrast, an additive modulation increases the response variance without altering slope, and thus information decreases. For instance, the neuron in the second panel of Fig 3A had a pure multiplicative gain, and therefore its responses to different orientations became more distinct with increasing population activity, potentially increasing the sensory information encoded. In contrast, the neuron in the first panel had also a large additive modulation, which could result in a drop in the information it encodes (since the response variance will be higher for the stronger responses). Thus, how information is affected by fluctuations in population activity will depend in part on the relative prevalence of multiplicative and additive modulation in the ensemble.

**The information encoded by neurons depends on the strength of population activity**

We tested the predictions of our model with our data. As an illustration, we first selected an orientation-selective neuron that had a strong multiplicative factor (1.8, permutation test $p < 0.002$; Fig. 5). The prediction of the multi-gain model is that the information encoded by this neuron about stimulus orientation should increase with population activity. As a proxy for information we used the decoding performance (fraction of correctly predicted stimulus orientation) of a multivariate logistic regression decoder (Experimental Procedures), cross-validated on hold-out trials that were not used to train the decoder. Better decoding performance corresponds to an increase in sensory information (Moreno-Bote et al., 2014). Although our decoder was trained on all orientations simultaneously, we split the performance into each orientation and obtained a separate decoding performance per orientation, as the non-uniformity of tuning curves caused some orientations to be better encoded than others. In addition, for each stimulus orientation we split the data into trials with either high or low population activity to characterize how population activity modulated information. When performing this analysis, we measured population activity as the summed activity of all recorded units, excluding the unit (or ensemble of units, see below) for which information was computed, just as when we characterized tuning modulation. For the selected unit, decoding performance increased substantially with population activity, by 44 and 9 percentage points for the two illustrated orientations (Fig. 5, top panel). This example shows that the sensory information encoded by neurons can vary substantially with the overall population activity.

**Sensory information in neuronal ensembles**

The multi-gain model predicts that neurons with stronger multiplicative effects should provide better performance with higher population activity than neurons with weaker multiplicative effects. Consistent with this prediction, we found a significant positive correlation across all units between the magnitude of the multiplicative factor and the performance change when moving from low to high population activity (Fig. 6A, left panel; $\rho = 0.6$, t-test, $p < 10^{-28}$; analysis based on responses measured 160-260 ms after stimulus onset). As also predicted by the multi-gain model, units with stronger additive effects had a larger negative performance change (Fig. 6A, right panel; $\rho = -0.32$, t-test, $p = 3\ 10^{-7}$). In summary, units with strong multiplicative effects provide more information as population activity increases, whereas units with additive modulation provide less information.

The multi-gain model also predicts that the balance of multiplicative and additive factors should determine how the information encoded by small neuronal ensembles, not just by units, should vary with the population activity level. This is true if responses are conditionally independent given the global modulatory factor (Eq. (1)) such that information in the ensemble becomes the sum of the "units" contributions (Eq. (2)). We tested this prediction by grouping orientation-selective units, including both single neurons and multiunits, in ensembles of size $N$ ($N = 1, 2, 3, 5, 10$ and $15$) as follows. Within each dataset, we ordered orientation-selective units by their multiplicative (or additive) factors, and then split them into non-overlapping ensembles of $N$ units that preserved this ordering (Experimental Procedures). For all datasets, the correlation between performance change and average multiplicative factor of the ensemble was positive and increased rapidly for larger sizes $N$ of the ensemble (Fig. 6B, left panel). Examples of these correlations are shown in Fig. 6C for ensembles of size $N=5$ (except D5, where individual units are shown). The datasets from anesthetized animals featured a strong positive correlation between performance change and the average multiplicative factor (Pearson's $\rho = 0.92$, $p = 7\ 10^{-6}$, D1; $\rho = 0.80$, $p = 4\ 10^{-5}$, D2; $\rho = 0.87$, $p = 0.002$, D3; $\rho = 0.78$, $p = 4\ 10^{-4}$, D4). The dataset from an awake animal (Fig. 6C, last panel) showed the same trend but did not reach significance (Pearson's $\rho = 0.66$, $p = 0.08$), most likely due to the small number of orientation-selective units available (8 neurons). On average across datasets, the performance change was roughly 10 percentage points for the ensembles of $N=5$ neurons with strongest multiplicative modulation, from a baseline performance of 44% correct (where chance performance is 12.5% in anesthetized data).

In contrast, the correlation between performance change and the average additive factor in the ensemble showed the opposite pattern: performance change was typically more negative for larger ensembles (Fig 6B, right). Examples of these negative correlations are shown in Fig 6D, following the same conventions as Fig. 6C. In 3 out of 4 anesthetized datasets, the correlation between performance change and additive factors was significantly negative (Pearson's $\rho = -0.48$, $p = 0.09$, D1; $\rho = -0.63$, $p = 0.003$, D2; $\rho = -0.97$, $p = 2 \cdot 10^{-5}$, D3; $\rho = -0.54$, $p = 0.03$, D4), while in the awake dataset the correlation was negative but not significant ($\rho = -0.58$, $p = 0.1$, D5). On average across datasets, the performance change was roughly -2% percentage points for the ensembles of $N=5$ neurons with strongest additive modulation.

When, instead of using population activity, we repeated the analysis described above with the projection of the population activity vector onto the first PCA component, we again found that information was differentially modulated in ensembles with strong multiplicative and additive effects (Figure S4C-F). We found similar but weaker results when information in the evoked response was conditioned on the strength of population activity measure just before stimulus onset (Figure S7), consistent with the modulation of tuning with pre-stimulus activity described in Fig. 4.

**Population activity does not substantially change total information, but redirects information into additively- and multiplicatively-modulated neuronal ensembles**

Thus far we have shown that information increases for multiplicatively-modulated ensembles and decreases for additively-modulated ensembles, when population activity is stronger. But what is the net dependence of information on the strength of population activity? To address this question, we randomly selected units to form neuronal ensembles of varying sizes ($N = 1, 2, 3, 5, 10$ and $15$), instead of choosing subsets of neurons based on their modulation as we did previously. We computed the performance change between low and high population activity, averaged across many ensembles. We found that there was little change in performance on average (Fig. 7A, black line). However, when we selected from these randomly-generated ensembles the 10% of cases with the strongest overall multiplicative modulation, we found that performance change was large and saturated as a function of ensemble size (green line), consistent with our previous analysis. Similarly, when we selected the 10% of ensembles with the strongest additive modulation, we found that performance change was consistently negative (blue line).

These results suggest that in randomly sampled neuronal ensembles, the effect of population activity on information is negligible. In fact, when we computed decoding performance in these populations at low and high population activity we did not find a visible modulation (Fig. 7B; the two lines overlay). Therefore, population activity does not substantially modulate the information present in these populations, but rather it modulates which ensembles have more information about the stimulus at different times: when population activity is high, the information encoded by multiplicatively-modulated ensembles is enhanced; when population activity is low, the information provided by additively-modulated ensembles is more important.

**Discussion**

We found that intrinsic fluctuations of stimulus-evoked and ongoing population activity are associated with multiplicative and additive modulation of the tuning of orientation-selective neurons in monkey V1. Neurons that showed strong multiplicative modulation tended to display weak additive modulation, and vice versa. These forms of modulation affected the sensory information encoded by neurons and small neuronal ensembles. As predicted by a multi-gain model, we found that sensory information increased with population activity for neuronal ensembles with strong multiplicative gains. However, sensory information decreased with greater population activity for ensembles with strong additive modulation. Importantly, we found that these effects largely offset each other, so that intrinsic fluctuations of population activity do not strongly affect total sensory information. Rather, the strength of population activity seems to act as a 'traffic light' that differentially redirects information into different subsets of neurons.

Previous work (Arieli et al., 1996; Tsodyks et al., 1999) found that pre-stimulus ongoing population activity has an additive effect on evoked responses, and others have reported either additive (Ecker et al., 2014) or multiplicative (Goris et al., 2014) modulations of evoked responses with population activity fluctuations. Recent work has shown that both multiplicative and additive modulations are present in mice and cat neuronal populations (Lin et al., 2015) and reported that a model with a single multiplicative factor across all neurons and different additive factors is favored. Instead, by observing how population activity affects tuning curves, we found that separate multiplicative and additive factors per neuron are required to describe our monkey data, a result further supported by a model comparison analysis (Figure S5). Importantly, we found that multiplicative and additive effects are not randomly intermixed across neurons. Rather, neurons with strong modulation of one type tend to show weak modulation of the other. In addition, we found that the strength of spontaneous activity preceding stimulus onset induces not only an additive modulation of stimulus-evoked responses as described in previous work (Arieli et al., 1996), but also a multiplicative effect. However, the influence of pre-stimulus population activity on tuning was weaker than that of stimulus-evoked population activity fluctuations, presumably because activity fluctuations have a timescale of a few hundreds of milliseconds.

Several recent studies have addressed how network state affects sensory responses and encoding, generally defining states based on the degree to which activity is synchronized across neurons or based on LFP measurements (Luczak et al., 2013; Mochol et al., 2015; Pachitariu et al., 2015; Scholvinck et al., 2015). In these works it is often reported that correlations are coupled with population activity measurements (Mochol et al., 2015; Pachitariu et al., 2015; Scholvinck et al., 2015). Much less attention has been paid, however, to the question of how within-state, across-trials fluctuations in the strength of population activity affect neuronal tuning and encoded information, although these fluctuations have been well-documented (Arieli et al., 1996; Kenet et al., 2003; Tsodyks et al., 1999). In our data, fluctuations of population activity strength do not correspond to changes in the degree of network synchronization, because neither variability nor correlations change substantially when going from low to high population activity (Figure S6). This may be because fluctuations of population activity in our data correspond to within-state fluctuations, rather than to across-state fluctuations. In fact, our analysis shows that the distributions of population activity are unimodal, suggesting a single state (Figure 2A). Overall, by performing an analysis in which population activity was the central quantity to condition on, we were able to reveal that across-trials fluctuations in the strength of population activity affect sensory tuning and the information encoded in distinct subsets of neurons.

Interestingly, our results show that the multiplicative effects on orientation tuning are as large in the awake animal as in the anesthetized preparation. This similarity in modulation occurred despite differences in the magnitude of pairwise correlations between our datasets from awake and anesthetized animals (median pair-wise spike count correlations in 160-260 ms window for anesthetized data: $\rho = 0.073$, D1; $\rho = 0.091$, D2; $\rho = 0.043$, D3; $\rho = 0.061$, D4; and for awake: $\rho = 0.013$, D5). However, the smaller pairwise correlations observed in the awake preparation nevertheless involved substantial shared fluctuation in the full population, which were clearly evident when we conditioned on the population activity of ~20 units. Therefore, although the magnitude of pairwise correlations might vary across experimental preparations (e.g. brain state, cortical areas, layers, etc.) (Cohen and Kohn, 2011; Ecker et al., 2014; Ecker et al., 2010; Kohn and Smith, 2005), their net effect on the population can be similar. Indeed, recent work has emphasized that the magnitude of pairwise correlations is not informative about their functional impact: even tiny correlations of a particular form called differential correlations can have massive effects on population information, whereas large correlations with a different structure can have little effect (Moreno-Bote et al., 2014).

A modulation of sensory tuning similar to the one that we report has been observed with optogenetic stimulation of specific V1 neuronal subpopulations. Optogenetic stimulation of layer 6 in mouse primary visual cortex induces divisive (i.e. multiplicative) gain modulation of orientation-selective neurons in the upper layers (Olsen et al., 2011). Similarly, optogenetic stimulation of inhibitory neurons in rat primary visual cortex has been shown to cause divisive or subtractive changes in the tuning of target neurons, depending on the inhibitory subpopulation that is stimulated (Wilson et al., 2012). More recently, antidromic spikes generated by optogenetic stimulation of distal V1 locations have been shown to additively and divisively modulate layer 2/3 neuronal responses in the mouse (Sato et al., 2014). These effects are similar to those we report, although future work will need to determine whether they provide a mechanistic explanation for the effects we observe under stimulus-driven conditions. An alternative explanation is that additive and multiplicative modulation can arise from balanced excitatory and inhibitory inputs (Chance et al., 2002). Specifically, multiplicative modulation arises from excitatory and inhibitory currents that are tightly balanced, whereas additive modulation might involve a slight imbalance in these currents. In this context, our results suggest that the balance of excitation and inhibition varies across neurons.

One might be tempted to equate the modulation of single neuron activity that we observe to that induced by the allocation of attention. Indeed, attention has been shown to modulate tuning in multiplicative and additive manners, similar to the modulation of tuning that what we have observed with population activity (Baruni et al., 2015; McAdams and Maunsell, 1999; Thiele et al., 2009; Treue and Martinez Trujillo, 1999). However, attention has been also shown to reduce response variability and pairwise correlations (Cohen and Maunsell, 2009; Mitchell et al., 2009) (but see (Ruff and Cohen, 2014)), whereas we found little change in these measures with fluctuations in population activity (Figure S6). Multiplicative modulation of tuning is also evident with manipulations of stimulus contrast (Carandini and Heeger, 1994; Finn et al., 2007; Priebe and Ferster, 2012). It is possible that the multiplicative modulation of tuning we report here shares similar mechanisms to those that occur with manipulations of stimulus contrast. In this regard, it is worth noting that the similarity of the multiplicative modulation we report to variations caused by altering contrast suggests that fluctuations in population activity limit information about stimulus contrast in V1, perhaps explaining limitations on perceptual contrast discriminability. This is because fluctuations that are identical to those generated by stimulus variations are the ones that limit information about the stimulus (Moreno-Bote et al., 2014).

The neuron-specific modulation of tuning with population activity fluctuations that we have characterized might govern important aspects of sensory processing, as these fluctuations affect the amount of sensory information that can be read out from small neuronal ensembles. For instance, the modulation might contribute to the trafficking of information in primary visual cortex, because an increase in overall activity tends to boost information in multiplicatively-modulated neurons while impoverishing it in additively-modulated neurons. Although we have shown that population activity does not substantially change total information in the recorded population, population activity through its neuron-specific multiplicative and additive modulations may act as a global context or 'traffic light' which influences which neuronal ensembles convey more information about the stimulus. In a speculative vein, efficient synaptic plasticity in small neuronal assemblies requires that their responses carry information about relevant internal and external variables (Fusi et al., 2007; Urbanczik and Senn, 2009), so modulating their information about those variables can also gate plasticity. Therefore, population activity might also control important aspects of learning.

# Experimental Procedures

## Animal preparation

We recorded data from five adult male macaque monkeys (*Macaca fascicularis*), four anesthetized and one performing a fixation task. The techniques used in anesthetized animals have been previously described (Smith and Kohn, 2008). Briefly, anesthesia was induced with ketamine (10 mg/kg) and maintained during preparatory surgery with isoflurane (1.5–2.5% in 95% O2). Sufentanil citrate (6–24 $\mu g$/kg/h, adjusted as needed for each animal) was used to maintain anesthesia during recordings (see Supplemental Experimental Procedures).

For experiments involving the awake monkey, the animal was implanted with a head post and then trained to fixate in a 1 deg window. Eye position was monitored with a high-speed infrared camera (Eyelink, 1000 Hz). 500 ms after the establishment of fixation, a drifting grating appeared over the aggregate receptive field of the recorded units. If the animal broke fixation, the trial was aborted and the data discarded. The animal was rewarded with a drop of water for successfully completed trials, typically ~500-800 per session.

All procedures were approved by the Institutional Animal Care and Use Committee of the Albert Einstein College of Medicine at Yeshiva University and were in compliance with the guidelines set forth in the United States Public Health Service Guide for the Care and Use of Laboratory Animals.

## Visual stimuli

For anesthetized animals, we presented full contrast drifting sinusoidal grating for 1280 ms, with an interstimulus period of 1500 ms. Gratings of 8 different orientation were each shown 300-400 times. For the awake animal, we used 12 different orientations and each was presented for 350 ms, with an interstimulus period of 50 ms. After every 4 stimuli, the monkey was rewarded. We recorded 50 trials for each stimulus orientation (see Supplemental Experimental Procedures).

## Recording methods and data preprocessing

We recorded in the superficial layers of primary visual cortex (V1), using a Utah array (96 microelectrodes, 1 mm length, 400 $\mu m$ spacing; 48 electrodes in the awake animal). Events crossing a user-defined threshold were digitized (30 kHz), saved, and sorted offline. We quantified spike waveform quality using a simple signal-to-noise ratio metric (SNR; (Kelly et al., 2007)). We defined multiunits (*multi-unit activity*; MUA) to be units with SNR>2, which corresponds to clusters with a small number of single units. Well-isolated units were defined to have SNR>3.5, a conservative threshold. A small number of MUA sites with SNR < 2 were also used only to compute the population activity.

We only analyzed blocks of trials in which responses did not change markedly over time. While the data from anesthetized animals were quite stable, those from the awake preparation showed strong evidence of adaptation. We therefore removed from this dataset the first 200 trials, leaving 400 trials during which responses were stable. To avoid biases due to a different number of trials per orientation in the decoding performance analysis, we (randomly) selected 30 trials for each orientation. Due to the low number of trials compared to the 'anesthetized' datasets, we merged trials with adjacent orientations in pairs to obtain 6 orientations with 60 trials per orientation to offer more comparable results. For all datasets, we measured responses beginning 60 ms after stimulus onset to account for V1 response latencies, as in (Graf et al., 2011).

**Dependence of tuning curves on population activity**

Tuning conditioned to population activity was computed for every orientation-selective unit using the following model-free approach. Orientation-selective neurons were defined as those with tuning well fitted by a von Misses function ($r^2 \geq 0.75$) (Graf et al., 2011); remaining neurons were termed non-selective. For each trial, we computed the population activity as the average number of spikes per second across all other neurons in that time window. Population activity was based on *all* neurons except the one for which tuning modulation was computed. Trials corresponding to a given stimulus orientation were sorted as a function of mean population activity, and then split at their median into subgroups of 'low' and 'high' activity. We computed the 'high' and 'low' population activity tuning, denoted $f^{high}(\theta)$ and $f^{low}(\theta)$, as the firing rate of the chosen neuron as a function of stimulus orientation in trial subgroups with 'high' and 'low' population activity, respectively. For comparison, we also computed population activity after z-scoring the responses of each neuron across trials. This approach yielded similar results, because trial ranking was nearly identical with the two methods, with only a few close-to-median trials changing category.

**Multiplicative and additive modulation of tuning**

To estimate the multiplicative and additive gains, we performed a type II weighted linear regression between the 'low' and 'high' tuning, using the model $f^{high}(\theta) = g\, f^{low}(\theta) + s$, where $g$ is the *multiplicative factor*, and $s$ is an additive offset. The multiplicative factor is unit-less by definition. To obtain a comparable unit-less *additive factor*, we normalized $s$ by the mean activity across orientations. Neurons with fit values outside the range [0.3,3] for multiplicative factors and [-1,1] for additive factors were excluded from analysis. Results do not qualitatively depend on the exclusion of these few outliers (5% of cases).

For each dataset, we estimated the gains' significance by a permutation test that sampled the null hypothesis. We randomly assigned trials to build 'high' and 'low' tuning, instead of ranking trials by population activity. Then, we obtained the multiplicative and additive factors for each neuron by linear regression, as described before, and computed the median factors across all neurons. We repeated this procedure 1000 times. We defined the probability that these medians were larger or smaller than the real median across neurons by the fraction of samples below or above the real population median. The reported two-tailed p-values were twice that fraction.

The statistical analysis for the correlation between multiplicative and additive factors (Fig. 3D) is described in the Supplemental Experimental Procedures.

**Broadening and displacement of tuning curves**

We determined the change in width and preferred orientation with population activity using Von Mises function fits to each neuron: $f(\theta) = a + b * \exp[k * [\cos(2 * (\theta - \theta_{pref})) - 1]]$, for trials with 'low' and 'high' population activity. We fit the function by minimizing the weighted squared error with bounded parameters to ensure physiologically plausible parameters (minimize function from lmfit python package with the following constraints: $\theta \in [0, \pi]$, $k > 0.001$, $\max(f_\theta) < 1.3 * \max(\bar{r_\theta})$, $\min(f_\theta) > 0.7 * \min(\bar{r_\theta})$, where $\bar{r_\theta}$ is the mean response across trials for each orientation) . To evaluate broadening, we used the squared width of the von Mises distribution, defined as $\sigma^2 = 1 - I_1(k)/I_0(k)$, where $I_n(k)$ is the modified Bessel function of the first kind of order *n* evaluated at *k*. The *broadening factor* was computed as the ratio between the widths of the 'high' and 'low' tuning ($\sigma^{high}/\sigma^{low}$). The *displacement* of the tuning was defined as the absolute difference between the preferred orientation of the 'high' and 'low' tuning ($\Delta\theta_{pref} = |\theta_{pref}^{high} - \theta_{pref}^{low}|$). We tested for significant broadening and displacement with a permutation test by sampling, as described above, using two-tailed and one-tailed p-values respectively. In an additional analysis, we used the von Mises fits to compute the multiplicative and additive modulation of tuning, which yielded similar results to those reported in the main text.

**Extracting visual information from population recordings**

We defined decoding performance (Fig. 5-7) as the fraction of trials where stimulus orientation was correctly predicted by a trained decoder. Results shown are for multinomial logistic regression (MLR) (Bishop, 2006) (see Supplemental Experimental Procedures), which outperformed a linear SVM decoder (an obvious alternative). We used 10-fold cross-validation (CV) to avoid over fitting the data; reported performance is the average performance across the 10 sets of left-out data.

**Decoding performance as a function of population activity**

We computed the decoding performance in the time window from 160-260 ms after stimulus onset using simultaneously recorded small ensembles of *N*=1, 2, 3, 5, 10, or 15 neurons. Decoding performance was independently analyzed for each orientation and each ensemble (Fig. 7) using MLR with the population vector formed by the firing rates of the neurons of the ensemble. Trials for each orientation were divided into 'low' and 'high' population activity trials, and their averages across trials were computed (Fig. 5). As for the tuning analysis, population activity was computed using all neurons but excluding the *N* neurons of the ensemble. Performance change per orientation was defined as the difference in decoding performance between high and low population activity trials for the MLR decoder trained in the two conditions and across orientations. We report the Pearson correlation coefficient and its two-tailed p-value, and we also plot a linear regression to highlight the relationship between performance change and mean factors in the ensemble.

**Figures and legends**

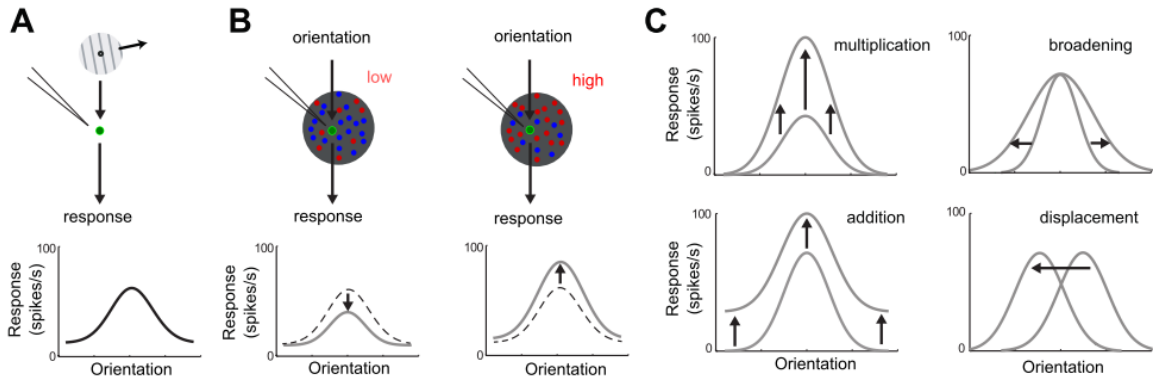

**Figure 1**: Hypothetical modulation of sensory tuning with population activity. (**A**) The 'classical' tuning of a neuron, computed without conditioning on population activity. (**B**) The firing rate of the neuron can depend on population activity. When population activity is low, tuning could have a lower gain (gray line, left); when population activity is high, tuning might have a higher gain (right). If the activity of the neuron is independent of population activity, the tuning for the two cases would be identical to the 'classical' tuning (dashed lines). (**C**) Tuning can be modulated in several ways with fluctuations in population activity, including multiplicative and additive effects, or both, and broadening and displacement.

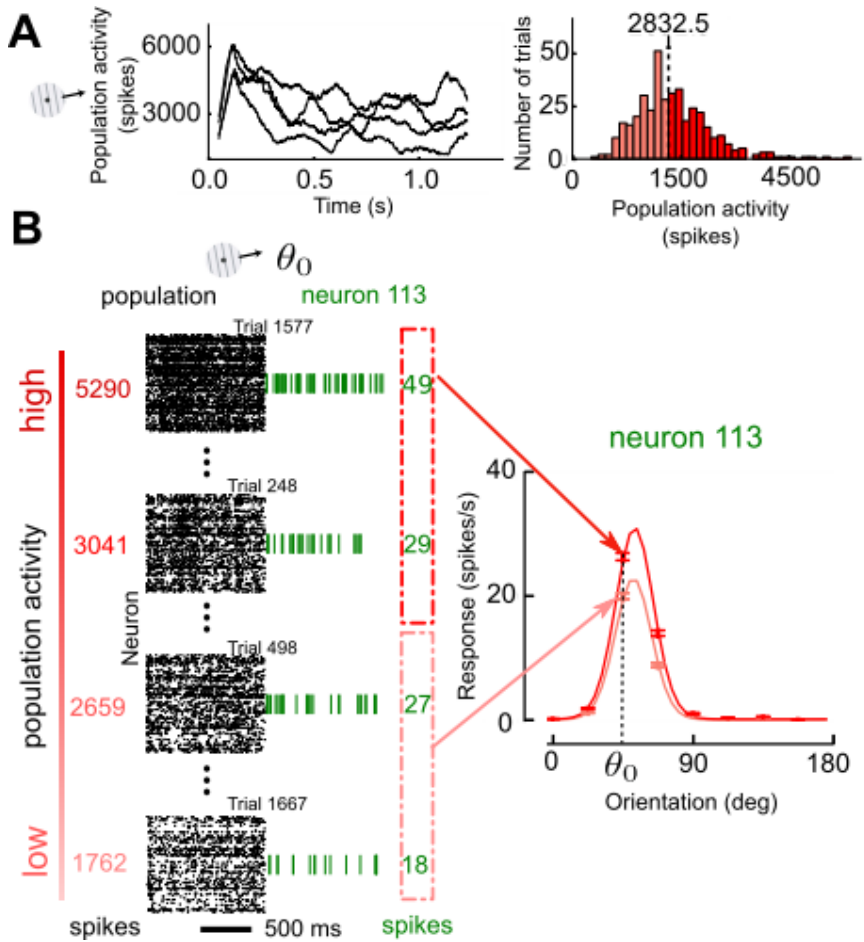

**Figure 2**: Sensory tuning depends on population activity in an example single neuron. (**A**) Stimulus-evoked population activity fluctuates across trials for the same stimulus (left panel). Four trials are shown. Distribution of population activity (sum of spikes across all neurons in the recorded population) across trials for a given stimulus (right). (**B**) The tuning of a single neuron (neuron 113 in D4) is strongly modulated with population activity (population activity is defined here as the sum of spikes across all neurons excluding the activity from the neuron for which tuning is being characterized). Population activity was ranked from high (top left) to low (bottom left) for each stimulus orientation $\theta_0$. The activity of the selected neuron (green spike trains) was averaged across either the top (red box) or bottom (light red box) 50th percentile of trials, and the averages were plotted as a function of stimulus orientation (rightmost panel). The tuning was modulated with population activity (red vs. pink lines), with stronger responses during periods of high population activity. Points and error bars are mean responses and s.e.m., respectively; lines are von Mises fits.

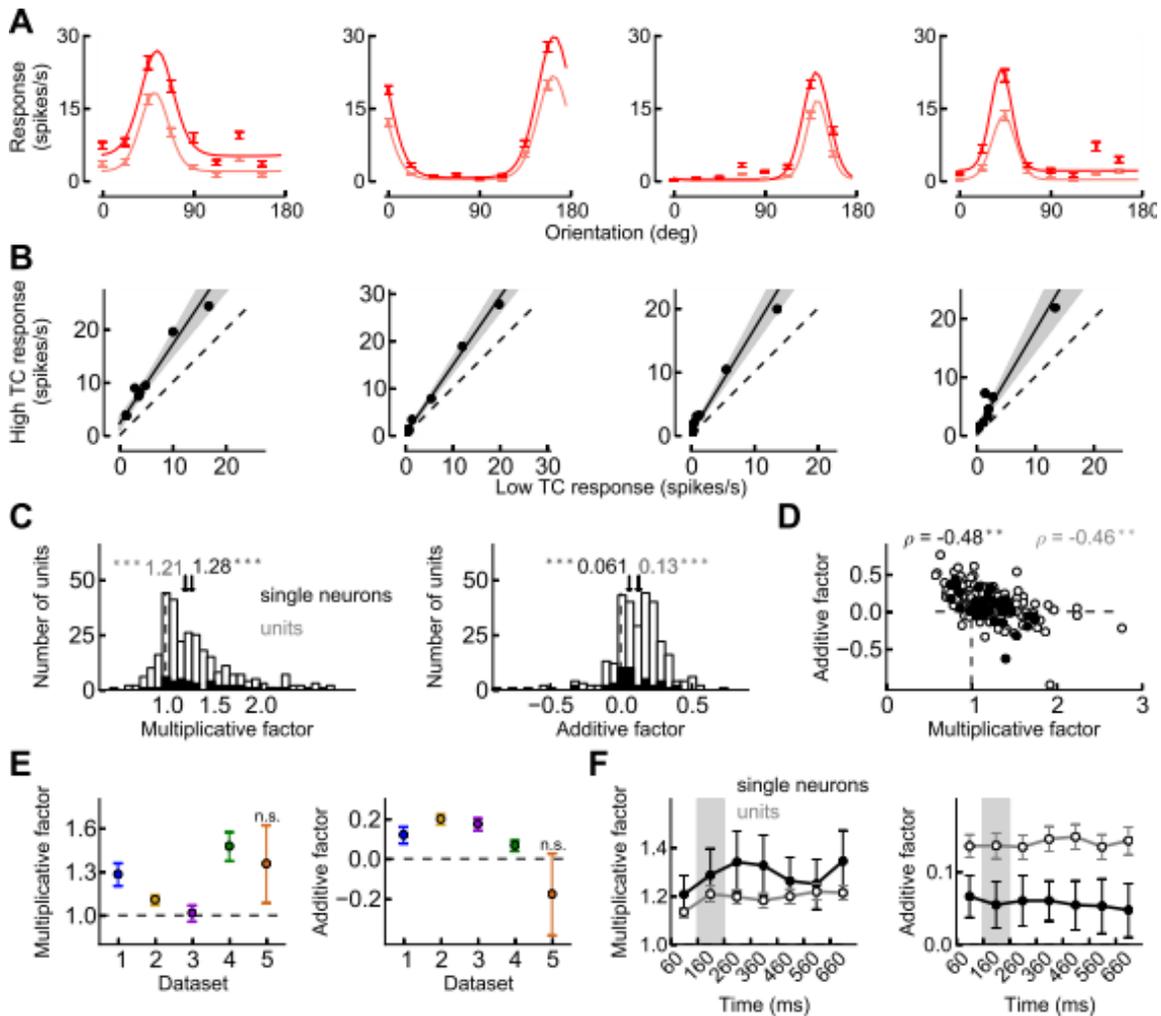

**Figure 3**: Sensory tuning undergoes multiplicative and additive modulation as a function of population activity. (**A**) Modulation of sensory tuning with population activity in four single neurons, computed as in Fig. 2B. Error bars indicate s.e.m. (**B**) Mean response of single neurons when population activity is high (ordinate, corresponding to red lines in (A)) vs. low (abscissa, light red lines in (A)). Each dot is the mean response to a different stimulus orientation. Shaded areas around the lines correspond to 95% confidence intervals. (**C**) Histograms of multiplicative (left panel) and additive factors (right) for all orientation-selective single neurons (black; $N = 45$) and orientation-selective units (white; $N = 293$). Median multiplicative factor across single neurons is 1.28, and across units is 1.21, shown in bold and non-bold formats respectively. Median additive factor across single neurons is 0.061, and across units is 0.13. (**D**) Additive and multiplicative factors

for both single neurons (black circles) and units (open circles; all) are negatively correlated. (**E**) For individual datasets, multiplicative factors (left panel) are typically significantly larger than one. For individual datasets, additive factors (right) are significantly larger than zero for all except for one dataset. (**F**) Median multiplicative and additive factors as a function of time (100ms time windows), relative to stimulus onset (time zero) across single neurons (black line) and units (gray). Shaded areas indicate time window (160-260ms) used to compute population activity and tuning curves in panels (A-E). Error bars correspond to 95% confidence intervals ($2.91 * m.a.d/\sqrt{N}$, where m.a.d is median absolute deviation). *: $p < 0.05$, **: $p < 0.01$, ***: $p < 0.001$, n.s.: $p > 0.05$.

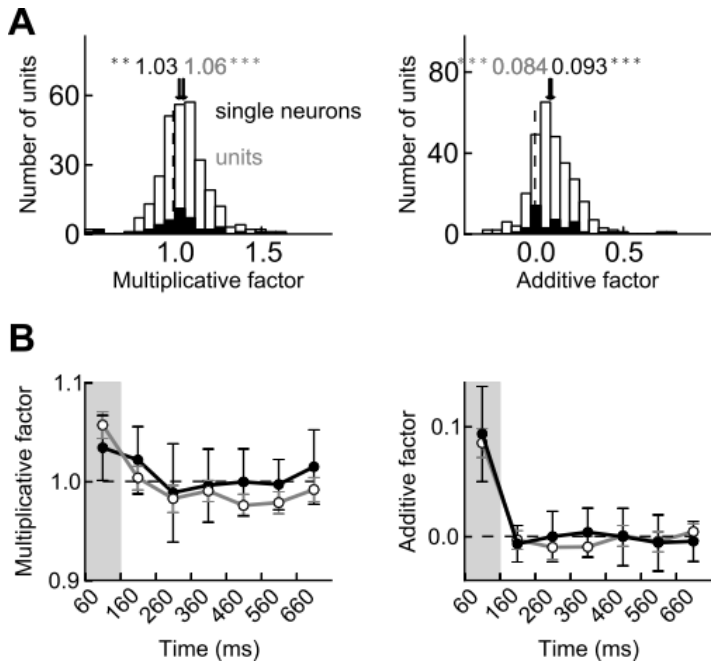

**Figure 4**: Sensory tuning depends on pre-stimulus population activity. (**A**) Histograms of multiplicative (left) and additive (right) factors across single neurons (black) and units (white). Median multiplicative factor across single neurons was 1.03 and 1.06 across units. Median additive factor across single neurons was 0.093 and 0.084 across units. (**B**) Median multiplicative and additive factors as a function of time (time windows of 100ms) for single neurons (black line) and all units (gray). Shaded areas indicate time window (160-260ms) used to compute statistics in panel (A), while population activity was computed during the pre-stimulus period (100ms before stimulus onset). Error bars indicate 95% confidence intervals ($2.91 * m.a.d./\sqrt{N}$). **: $p < 0.01$, ***: $p < 0.001$.

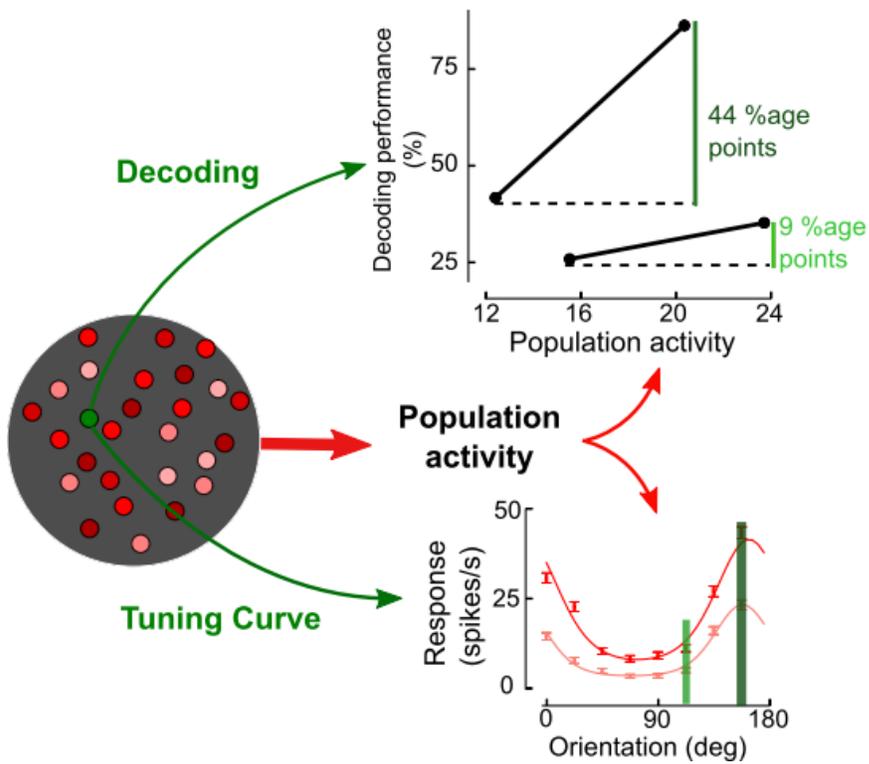

**Figure 5**: Information increases with population activity in a neuron with strong multiplicative modulation. This unit (from D1) had a large multiplicative factor (MF = 1.8, permutation test $p < 0.002$; AF = 0.16, $p < 0.002$). Decoding performance per orientation (cross-validated probability of correctly predicting the orientation on a trial by trial basis) for the selected unit increases with evoked population activity for two sample orientations (top right panel; performance changes are indicated for the two orientations).

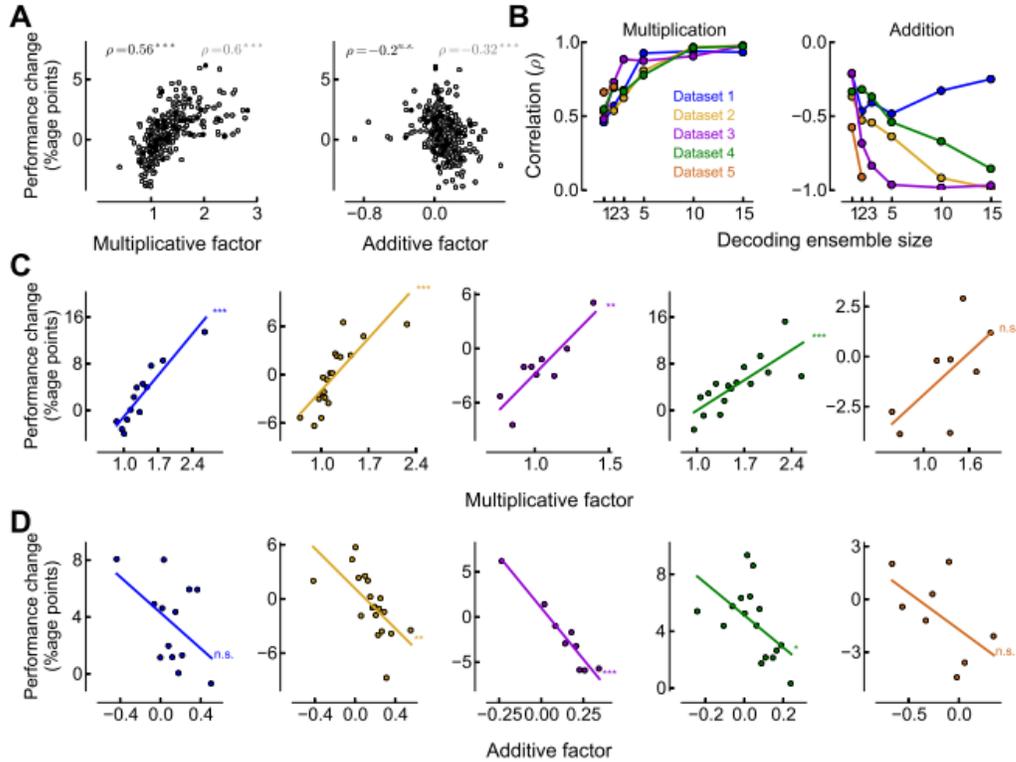

**Figure 6**: Performance change increases for units and ensembles of units with strong multiplicative modulation, and decreases for those with strong additive modulation when population activity is higher. (**A**) Performance change as a function of multiplicative (left) and additive (right) factors for all orientation-selective single neurons (black circles) and units (open circles; all). (**B**) Correlation between performance change and multiplicative (left) or additive factors (right), as a function of the number of units $N$ in the ensemble for each dataset (same color code as in Fig. 3E). Decoding is based on the entire ensemble, and the multiplicative and additive factors refer to the average factors in the ensemble. Population activity was measured after excluding the ensemble used to decode stimulus orientation. The correlation between performance change and average multiplicative factor in the ensemble increases with ensemble size and then asymptotes. In contrast, the correlation between performance change and average additive factor drops with ensemble size, as predicted by the multi-gain model. (**C**) Performance change increases with the strength of multiplicative modulation for each dataset individually (ensembles of $N=5$, except $N=1$ for D5). (**D**) Performance change decreases with the average additive factor of the ensemble (same sizes as panel C), and can even become negative. *: $p < 0.05$, **: $p < 0.01$, ***: $p < 0.001$, n.s.: $p > 0.05$.

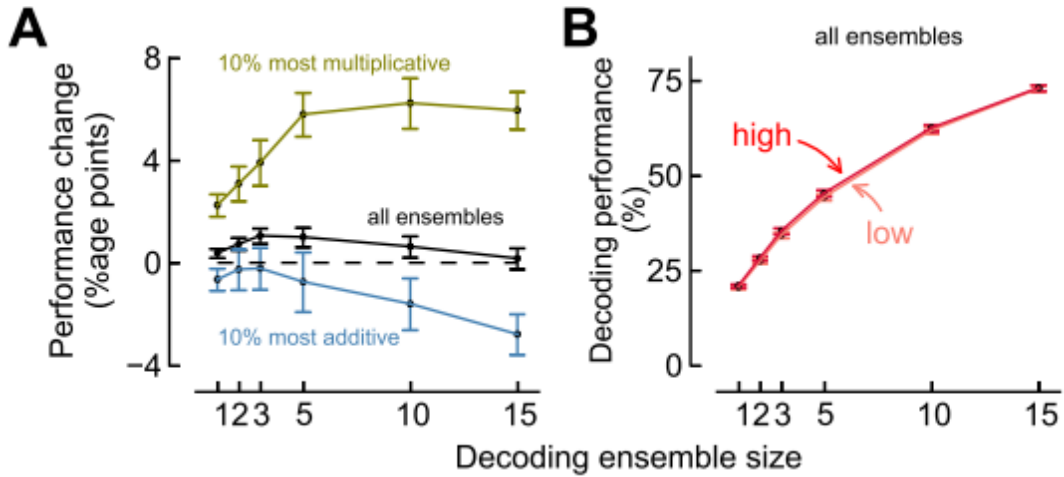

**Figure 7**: Population activity does not substantially change total information, but rather it differentially redirects information into multiplicatively and additively modulated neuronal ensembles. (**A**) Change in decoding performance averaged across randomly chosen ensembles of varying size (solid line), and for the top 10% ensembles with strongest average multiplicative (green) and additive (blue) factors. (**B**) Lack of modulation of performance with population activity as a function of ensemble size for randomly chosen ensembles.

# Supplemental Figures and Legends

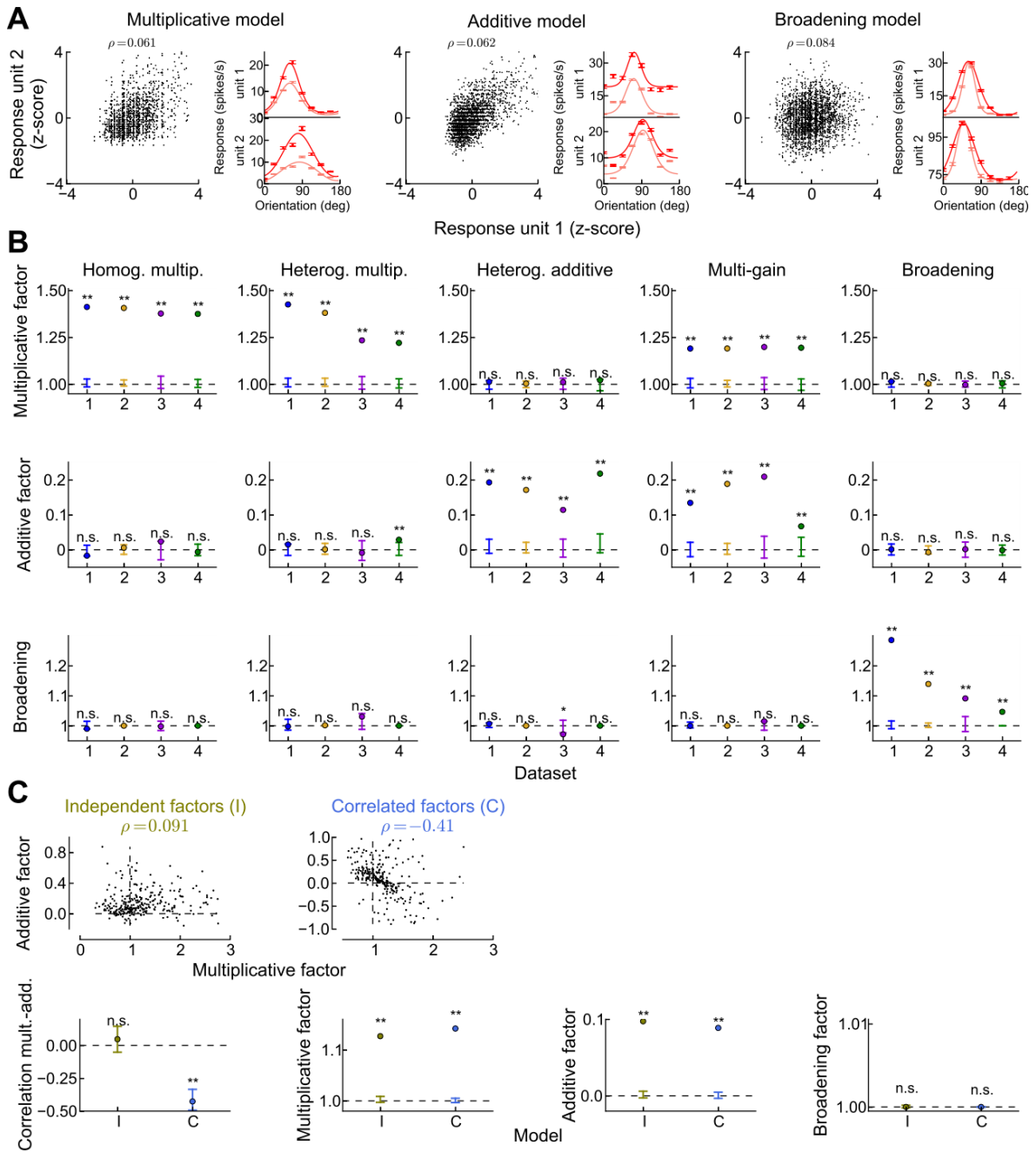

**Figure S1**: (Related to Figure 1) Positive correlations do not necessarily imply both additive and multiplicative effects, nor lack of broadening.

(**A**) Positive pair-wise correlations can be generated either with a purely multiplicative (left panel), a purely additive (center) or a broadening (right) model. These models generate positive correlations in the responses of pairs of neurons. The tuning curves of two simulated pairs for each model are modulated by population activity (small panels), like in the data. Across orientations, there is a correlation between their z-scored responses.

(**B**) Validation of the estimation method of multiplicative and additive factors.
The method is validated using simulations with neurons having tunings curves as in the data, and with similar average correlation coefficients. We manipulated the presence and the combination of contributing factors (multiplication, addition and broadening) to global fluctuations. The method reliably detects the presence of significant multiplicative, additive or broadening effects, and rejects their presence if they are absent. Each column provides simulations of neuronal populations with tuning curves as in datasets D1-D4, and the outcome of the estimation of multiplicative, additive and broadening factors for each case. The method reliable finds the factors that are present in the simulated data when the factors are substantial, and rejects the presence of factors that are not present in the data. For instance, we first simulated data using a model (the homogenous multiplicative model; first column) with purely multiplicative factors and no addition or broadening. Our estimation method, when applied to this simulated data, correctly discovers significant multiplicative factors (top panel) and correctly does not find either additive factors (middle) or broadening (bottom). We repeated this procedure with several combinations of factors to validate our estimation method in a broad set of conditions, always showing the general validity of our method (see details of the models next).

From left to right, we consider the following models: (i) homogeneous multiplicative model, with same gain for all neurons, (ii) heterogeneous multiplicative model, characterized by having a potentially different multiplicative factor for each neuron, (iii) heterogeneous additive model, with different additive factors for each neuron (iv) multi-gain model where multiplicative and additive factors are independently and heterogeneously assigned to the neurons in the populations, but with the same global modulation for both factors and (v) broadening model, characterized by having fluctuations of the width of the tuning curve and no other modulatory factor. Mathematically, these models are specified by the mean firing rate $f_i(\theta, g)$ for neuron $i$ given the stimulus $\theta$ and the modulatory global factor $g$. For each model, the mean firing rate is:
(i) Homogeneous multiplicative model: $f_i(\theta, g) = (1 + g)h_i(\theta)$ (based on Goris et al, 2014)
(ii) Heterogeneous multiplicative model: $f_i(\theta, g) = (1 + \alpha_i g)h_i(\theta)$
(iii) Heterogeneous additive model: $f_i(\theta, g) = h_i(\theta) + \beta_i g$ (following Lin et al, 2015)
(iv) Multiplicative and additive model: $f_i(\theta, g) = (1 + \alpha_i g)h_i(\theta) + \beta_i g$,
(v) Broadening model: $f_i(\theta, g) = h_i(\theta, g)$, where the global modulatory factor enters inside the tuning curve and rescales its width. Because the fluctuations of the width happen simultaneously across all neurons, this effect introduces purely positive pair-wise correlations (see panel A, right).

For each model, spike counts in a time window of 100ms (mimicking the time window 160-260ms used in the data analysis of Fig. 2) were drawn from an independent Poisson distribution across neurons with mean $f_i(\theta, g)$ as specified above. The global modulatory factor $g$ is shared among neurons, which introduced pair-wise correlations. The mean of the modulatory factor $g$ was zero. The values of $g$ were drawn from a gamma distribution with unit mean (the variance is specified below), followed by subtracting one, to keep $g$ bounded from below by -1 and its mean at zero. The parameters $\alpha_i$ and $\beta_i$, which control the heterogeneity in multiplicative and additive factors, were drawn from Gaussian distributions with mean and standard deviation respectively for each model (ii) $<\alpha_i> = 0.25$, $\sigma_\alpha = 0.25$, (iii) $<\beta_i> = 0.35$, $\sigma_\beta = 0.2$, and (iv) $<\alpha_i> = 0.13$, $\sigma_\alpha = 0.2$, $<\beta_i> = 0.8$, $\sigma_\beta = 0.8$. The parameters $\alpha_i$ and $\beta_i$ where drawn once per neuron, and did not subsequently vary across trials for a given model. The modulatory global factor, in contrast, was re-drawn for each trial, and so varied across trials. For all the models, the average tuning curves $h_i(\theta)$ constituted those fitted to real data. The variances ($\sigma_g^2$) of the global modulatory factor $g$ took the values (i) 0.05, (ii) 2.5, (iii) 2.5, (iv) 0.33, (v) 0.4, chosen to provide values of positive average correlations comparable to those in our datasets. The median correlation coefficient obtained in this 100 ms bin across datasets for each model was: (i) 0.07, (ii) 0.06, (iii) 0.08, (iv) 0.07, (v) 0.003, while in real data it was 0.07.

(**C**) The method reliably detects the presence of negative correlations between multiplicative and additive factors, and rejects its presence if such a correlation does not exist in the simulated data.
We computed the multiplication, addition, and the correlation between both factors for two models: one with correlations between multiplication and addition (C model), $f(\theta) = (1 + \alpha_i g)h_i(\theta) + \beta_i g$, where $\beta_i = b_i(c_i - \alpha_i)$, and parameters $b_i = 24$, $c_i = 0.5$, $<\alpha> = 0.26$, $\sigma_\alpha = 0.8$, $\sigma_g^2 = 0.05$; and another model with uncorrelated factors (I model), with different modulations for multiplication and addition $f(\theta) = (1 + \alpha_i g_1)h_i(\theta) + \beta_i g_2$, where $<\alpha> = 0.15$, $\sigma_\alpha = 0.5$, $<\beta> = 2.5$, $\sigma_\beta = 1$, $\sigma_{g_1}^2 = \sigma_{g_2}^2 = 0.5$. The pair-wise correlations are 0.16 and 0.18 for the C and I models, respectively, close to the values for the real data (0.21). These models were simulated using a longer time window to allow direct comparison with the data analysis performed in Fig. 3D. The correlations between multiplicative and additive factors was negative and significant for the model with negatively correlated factors ($\rho_C = -0.34$, non-parametric bootstrap p-value = 0.002) and not significant for the one with independent factors ($\rho_I = 0.048$, non-parametric bootstrap p-value = 0.4).

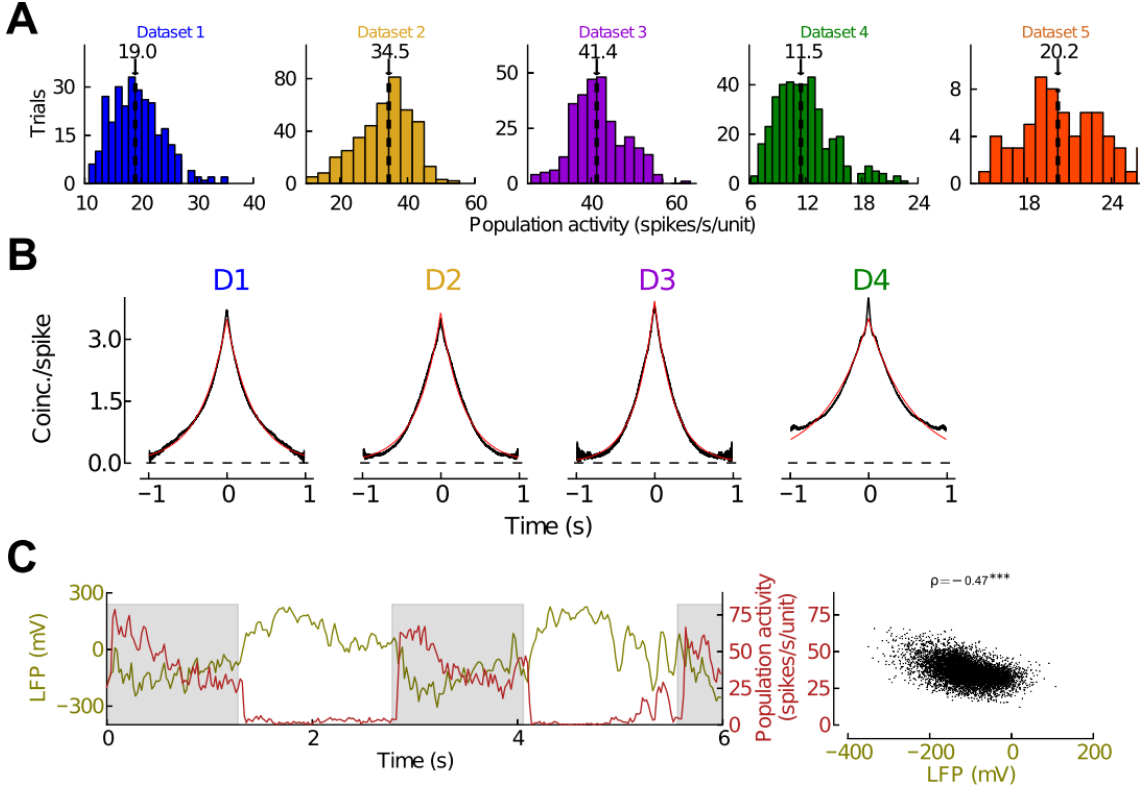

**Figure S2**: (Related to Figure 2) (**A**) Distributions of population activity for all datasets, computed as in Fig. 2A. The distributions tend to be unimodal and wide. (**B**) The autocorrelation of spontaneous population activity fluctuations is well-approximated by an exponential with a time constant of around 300 ms. We computed the autocorrelation (black lines) of population activity during the last second of spontaneous activity before stimulus onset, excluding the 500 ms period after stimulus offset to avoid transients. The autocorrelation was computed for trials with the same previous stimulating grating, and then averaged across stimulus orientations, following the method used by (Kohn and Smith, 2005):

$ACG_j(\tau) =$
$\frac{N_j}{N_j-1} * \frac{1}{(T-|\tau|)\bar{R}} \left( \frac{1}{N_j} \sum_{k}^{N_j} \sum_{t}^{T-\tau} R^k(t) R^k(t+\tau) - \frac{1}{N_j-1} \sum_{k}^{N_j-1} \sum_{t}^{T-\tau} R^k(t) R^{k+1}(t+\tau) \right)$,

where $R^k(t)$ is the population activity of trial k at time *t* (sum of spikes of orientation selective units in that millisecond), $\bar{R}$ is the mean population activity, j is the orientation, and $N_j$ is the number of trials with orientation j. The autocorrelations were well-fit by exponential functions (red lines) with time constant $\tau$ of: D1 337 ms, D2 300 ms, D3 251 ms, D4 546 ms. (**C**) Population activity is negatively correlated with the LFP. The left panel shows the mean LFP and population activity computed in time intervals of 20 ms across a few consecutive trials, with shaded area indicating the stimulus presentation periods. In the right panel the correlation between the LFP and the population activity computed in 6 time bins of 200 ms each during evoked activity across all orientations is shown. The observed correlation was negative and significant ($\rho = 0.47$, p-value $< 10^{-20}$, two-tailed t-test).

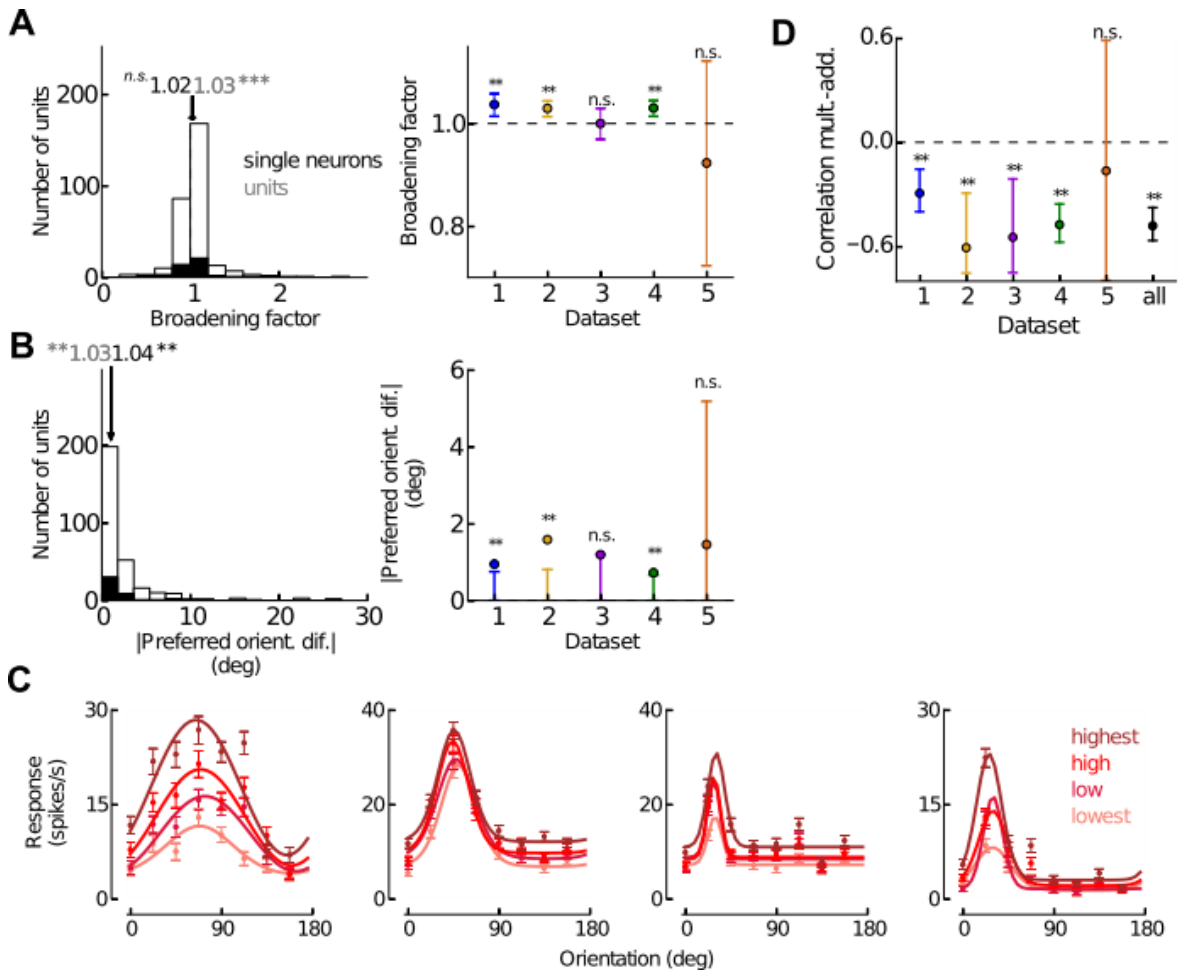

**Figure S3**: (Related to Figure 3) (**A**) and (**B**) Tuning width and preferred orientation are only weakly modulated with fluctuations in population activity. (**A**) Histogram (left) of the broadening factor (ratio of standard deviations of high and low population activity tuning curves) for all single neurons (black) and all units (white), and median values for each dataset (right), as obtained from von Mises fits to the tuning measured over the entire stimulus period (see Experimental Procedures). We found a small (2% for single neurons and 3% relative change for all units) widening of tuning, as a function of population activity (left panel; Mann-Whitney $U = 855$, $p = 0.2$, single neurons; $U = 3 \cdot 10^4$, $p < 10^{-11}$, all units). This effect was present in three datasets out of five (right, all units; median = 1.04, permutation test $p < 0.002$, D1; median = 1.04, $p < 0.002$, D2; median = 1.0, $p = 0.5$, D3; median = 1.03, $p < 0.002$, D4; median = 0.92, $p = 0.5$, D5). Error bars correspond to 95% confidence intervals. (**B**) Same as (A), but for the absolute difference in preferred orientation between the high and low activity tuning. Tuning preference was weakly but significantly modulated by population activity (left; median shift of 1 degree for single neurons and units, permutation test $p < 0.001$ in both cases). This effect was significant in 3 of the 5 datasets (right; median = 0.91, permutation test $p < 0.001$, D1; median = 1.6, $p < 0.001$, D2; median = 1.2, $p = 0.05$, D3; median = 0.75, $p$

< 0.001, D4; median = 1.4, $p = 0.3$, D5). Error bars show the 95% confidence intervals of the null hypothesis. *: $p < 0.05$, **: $p < 0.01$, ***: $p < 0.001$, n.s.: $p > 0.05$. (**C**) Tuning modulation as a function of population activity divided into four bins (25$^{th}$, 50$^{th}$, 75$^{th}$ and 100$^{th}$ percentiles; quartiles), ranging from low (lighter lines) to high population activity (darker lines). Population activity was computed in the 160-260 ms period. Error bars correspond to s.e.m and lines are von Mises fits. Same neurons as in Fig. 3A. (**D**) Correlation between multiplicative and additive factors dataset by dataset. The correlation is negative for all datasets, and significant in 4 out 5 datasets. Error bars show 95% confidence intervals computed by a non-parametric bootstrap test. *: $p < 0.05$, **: $p < 0.01$, ***: $p < 0.001$, n.s.: $p > 0.05$.

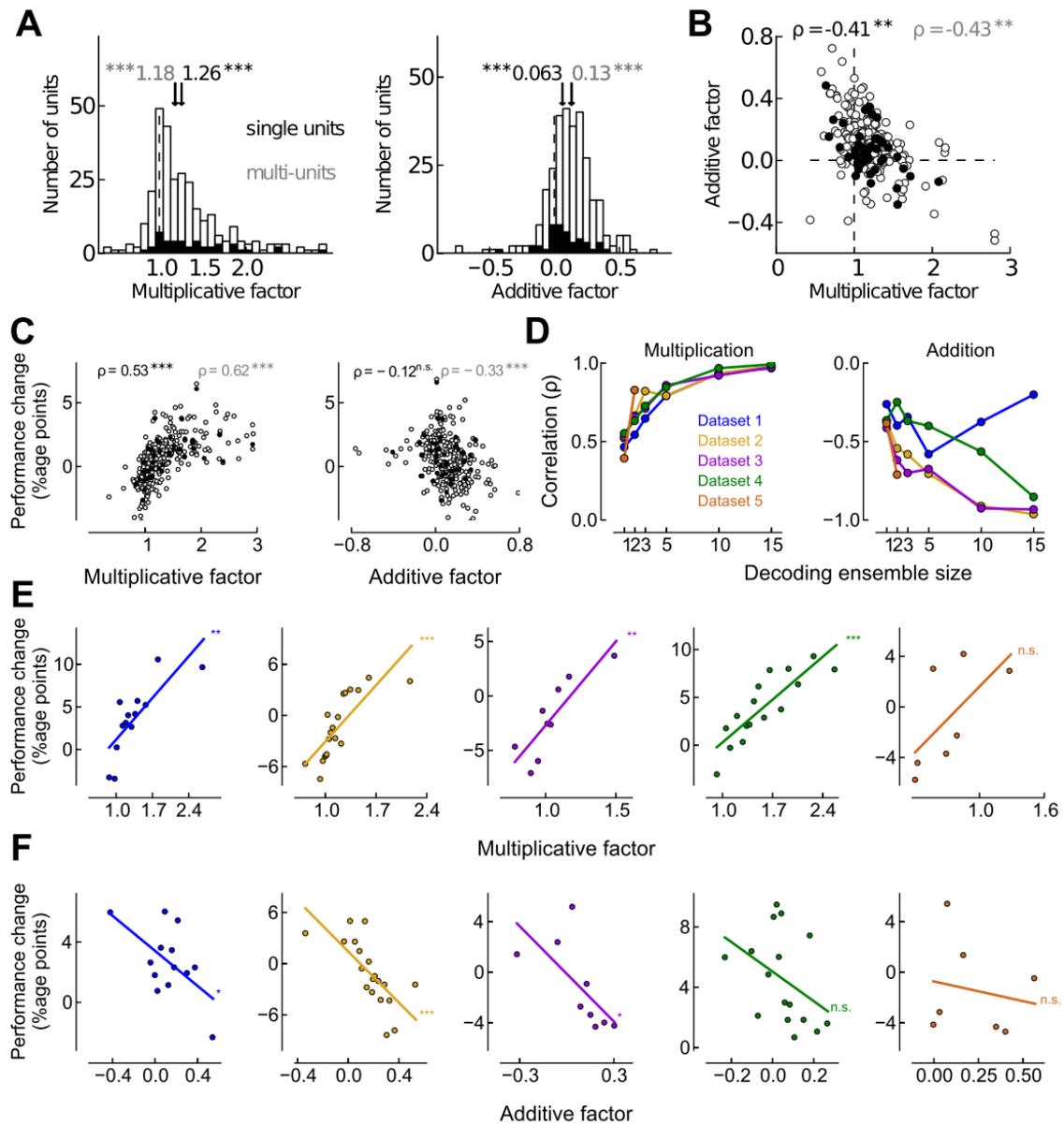

**Figure S4**: (Related to Figure 3) The projection of the evoked population activity vector onto the first principal component (PC) obtained by PCA modulates tuning curves and information much like population activity does. (**A**) The projection of population activity vector (vector whose elements are the spike counts of each neuron in the time period 160ms to 260ms) onto the first PC (vector capturing the largest amount of variance of the data, obtained by PCA) modulates V1 neurons both multiplicatively and additively. We applied PCA to a data matrix where each row corresponds to the population activity vector for each trial, and where the mean of each column has been subtracted. Factors were computed as in Fig. 2C while classifying trials as 'high' or 'low' based on the value of the projection of the population activity vector onto the first principal component. Color codes and time bins used are as in Fig. 2C. (**B**) The multiplicative and additive factors obtained using the first PC projection are negatively and significantly correlated. (**C**) Performance change from 'high' to 'low' activity (computed from the first PC

projection) as a function of multiplicative (left) and additive (right) factors (from first PC projection) for all orientation-selective single neurons (black circles) and units (open circles; all). (**D**) Correlation between performance change and multiplicative (left) or additive factors (right), as a function of the number of units $N$ in the ensemble for each dataset. (**E**) Performance change increases with the strength of multiplicative modulation for each dataset individually (ensembles of $N=5$, except $N=1$ for D5). (**F**) Performance change decreases with the average additive factor of the ensemble. *: $p < 0.05$, **: $p < 0.01$, ***: $p < 0.001$, n.s.: $p > 0.05$.

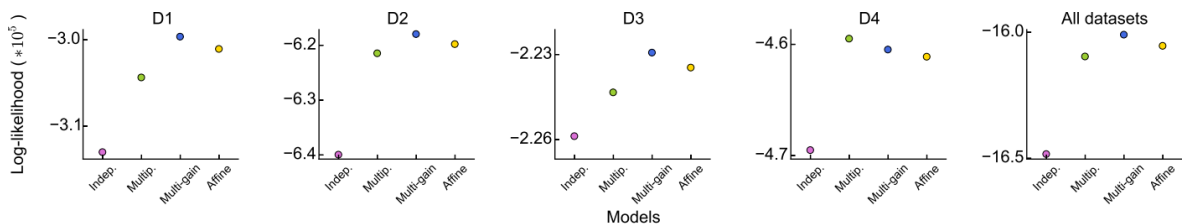

**Figure S5**: (Related to Figure 6) A model comparison between independent, homogenous multiplicative, multi-gain and affine models favors multi-gain models. The cross-validated log-likelihood of the data given the model is plotted individually for each simulated dataset (D1-D4, first 4 columns) and combined across datasets (last column). A higher log-likelihood means that the model better predicted the data of a hold-out set. As we use cross-validated log-likelihood (see details below), it is not necessary to take into account the different number of parameters for each model, as cross-validation already controls for model complexity. For each model, the gain-conditioned, Poisson firing rates were as follows:

(i) Independent model: $f_i(\theta) = h_i(\theta)$

(ii) Homogeneous multiplicative model $f_i(\theta, g) = (1 + g)h_i(\theta)$ (based on Goris et al, 2014)

(iii) Affine model: $f_i(\theta, g) = g_1 h_i(\theta) + \beta_i g_2$ (following Lin et al, 2015)

(iv) Multi-gain model: $f_i(\theta, g) = (1 + \alpha_i g)h_i(\theta) + \beta_i g$.

The models were trained on 80% of the data by optimizing model parameters to maximize likelihood of this training set. The reported log-likelihoods are for predicting the remaining 20% of the data, averaged over 10 independent splits of the data. This analysis was performed using the same time window of 100 ms (160-260 ms after stimulus onset) as in most of the other figures throughout the paper.

For the affine and multi-gain models the optimization procedure was as follows: first we maximized the log-likelihood of the model on the training set by coordinate ascent, maximizing the likelihood with respect to each of the parameters (the $\beta_i$-s, the $g_1$-s, and the $g_2$-s for the affine; and the $\alpha_i$-s, the $\beta_i$-s and $g$-s for the multi-gain model) alternatively while keeping the other parameters fixed. For the multi-gain model first we maximized the log-likelihood with the $\alpha_e$-s and $\beta_i$-s fixed, and obtained the $g$-s. Then, we maintained the $g$-s and the $\beta_i$-s constant while maximizing the log-likelihood to get the new $\alpha_i$-s. Finally, we maximized the likelihood again but allowing only the $\beta_i$-s to change. A similar procedure was follow for the affine model. Note that while in Fig. S1 the mean values of the modulatory factors $g$ were zero, in this analysis the mean values are fitted. We performed each maximization step using the Large-scale Bound-constrained Optimization (L-BFGS-B) method. Once we obtained these parameters, we computed for each neuron the modulatory factors ($g_1$-s, and the $g_2$-s for the affine, $g$-s for the multi-gain model) in the test trials, removing from the population in each case the neuron whose modulatory factor was being computed, and maximizing the likelihood of each trial with the Sequential Least SQuares Programming (SLSQP) method. Then, we computed the log-likelihood of the test set using the parameters from the train set, and the global modulatory factors from the test set.

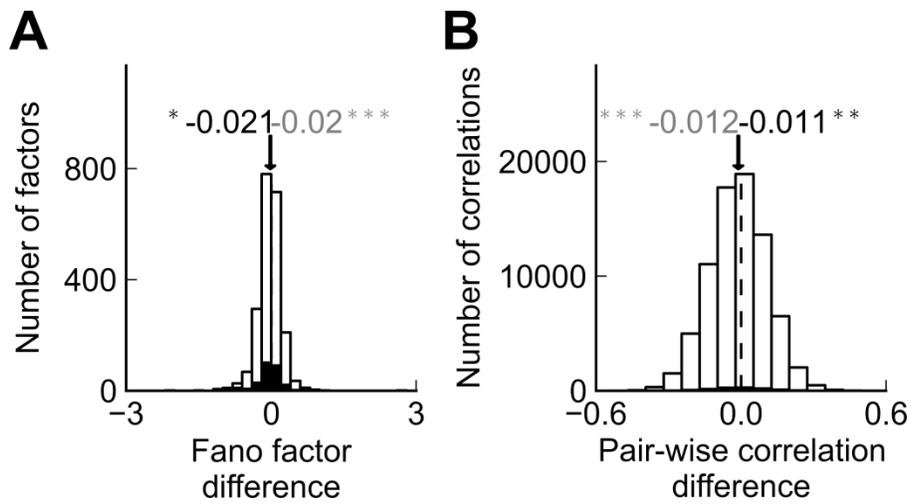

**Figure S6**: (Related to Figure 6) Fano factors of the spike counts and pairwise correlations depend weakly on population activity. (**A**) Median Fano factor was slightly but significantly smaller when population activity was high compared to when it was low (single units, in black: median = -0.021, Mann-Whitney $U = 3 \ 10^3$, $p = 0.02$, multi-units, in white: median = -0.020, $U = 2 \ 10^6$, $p < 10^{-4}$). (**B**) Same as before for the difference in spike count correlations (single units: median = -0.011, Mann-Whitney $U = 6 \ 10^5$, $p = 0.001$; multi-units: median = -0.012, $U = 3 \ 10^9$, $p < 10^{-4}$). *: $p < 0.05$, **: $p < 0.01$, ***: $p < 0.001$, n.s.: $p > 0.05$.

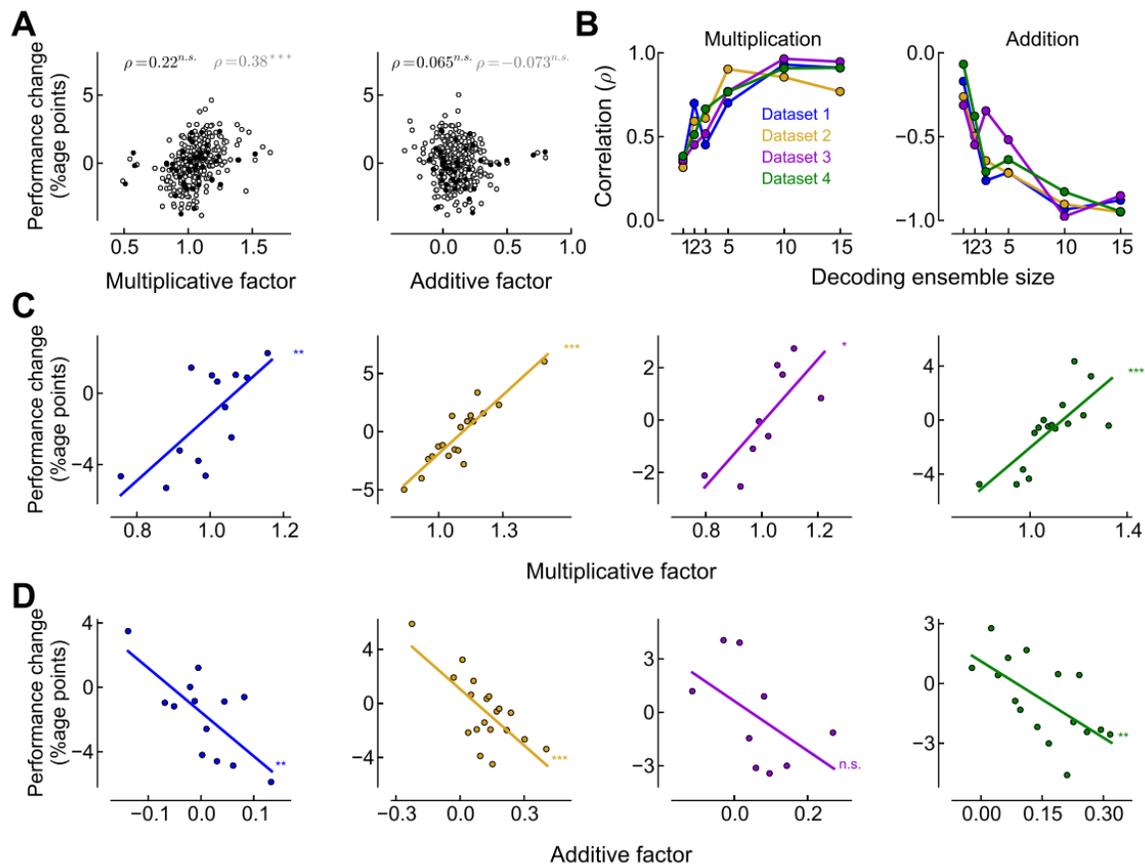

**Figure S7**: (Related to Figure 6) Performance increases for ensembles with strong multiplicative modulation and decreases for ensembles with strong additive modulation, when comparing trials with low versus high pre-stimulus population activity, much like for the same analysis using population activity measured during stimulus presentation. Panels are as in Fig. 6, except that awake data is not shown because of its short inter-stimulus intervals. Pre-stimulus population activity was measured in the 100 ms preceding stimulus onset, and performance changes were computed from 60 to 160 ms after stimulus onset.

# Supplemental Experimental Procedures

## Animal preparation

The techniques used in anesthetized animals have been described in detail previously (Smith and Kohn, 2008). Extending the explanations given in the main text, anesthesia was induced with ketamine (10 mg/kg) and maintained during preparatory surgery with isoflurane (1.5–2.5% in 95% O2). Sufentanil citrate (6–24 $\mu g$/kg/h, adjusted as needed for each animal) was used to maintain anesthesia during recordings. Eye movements were suppressed with vecuronium bromide (0.15 mg/kg/h). Drugs were administered in normosol with dextrose (2.5%) to maintain physiological ion balance. We monitored physiological signs (ECG, blood pressure, SpO2, end-tidal CO2, EEG, temperature, urinary output and osmolarity) to ensure adequate anesthesia and animal well-being. Temperature was maintained at 36–37 C°.

## Visual stimuli

Visual stimuli were presented on a CRT monitor, at a resolution of 1024 x 768 pixels and a video frame rate of 100 Hz. The display had a mean luminance of 40 cd/m$^2$, and was placed 110 cm away from the animal where it subtended 20º of visual angle. We generated stimuli with custom software based on OpenGL (EXPO). The spatial (1.3-2 cpd) and temporal frequency (6-6.25 Hz) of the gratings were chosen to correspond to the usual preference of parafoveal V1 neurons . Stimuli were shown in a circular aperture surrounded by a gray field of average luminance. Receptive fields in both awake and anesthetized animals were 2-4 degrees from the fovea. The size of the gratings (2-4 degrees in diameter) was chosen to cover the receptive fields of all the neurons.

## Multiplicative and additive modulation of tuning

To estimate the correlation between multiplicative and additive factors (Fig. 3D) from the linear fits described above we proceeded as follows. Because the linear model contains two regressors, corresponding to the multiplicative and additive factors, a linear fit based on the same data points can create artificial correlations between the values of these two regressors (Donahue and Lee, 2015). To avoid this artifact, we estimated the factors from randomly divided trials into two halves with the same number of trials for each orientation. A Pearson correlation coefficient sample was obtained by computing this coefficient between the multiplicative factors obtained from fits in the first trial subgroup and the additive factors obtained from fits in the second subgroup. Using two non-overlapping groups of trials ensured that the factors were not trivially correlated (Donahue and Lee, 2015). This procedure was repeated 1000 times (thus obtaining 1000 samples) to build a distribution of correlations. The reported correlation was the median across samples (the displayed distribution of multiplicative and additive factors in Fig. 3D is originated from a randomly chosen subdivision of the trials). The two-tailed p-value for the reported correlation was computed as twice the fraction of samples that are below or above zero, whichever is the smaller quantity. To improve the signal to noise ratio, the analysis used responses measured during the full trial (excluding the first 60 ms).

**Extracting visual information from population recordings**

We defined decoding performance (Fig. 5-7) as the fraction of trials where stimulus orientation was correctly predicted by the multinomial logistic regression (MLR) (Bishop, 2006). In brief, in MLR the probability of orientation $\theta_j$ is modeled as: $Pr(\theta_j|\mathbf{r}) = \exp(\mathbf{w}_j \cdot \mathbf{r}) / \sum_k \exp(\mathbf{w}_k \cdot \mathbf{r})$, where $\mathbf{r} = (r_1, r_2, ...)$, is a vector of the individual firing rates of the simultaneously recorded neurons, defined as the spike count in a given time window per trial divided by the time window: $r_i = counts_i/time$. The sets of vector parameters $\mathbf{w}_k$ are learned using maximum likelihood estimation using the function (*smlr*) from the PyMVPA Python package (Hanke et al., 2009). For each trial, the predicted stimulus orientation is given by $\theta_j^{pred} = argmax_j [Pr(\theta_j|\mathbf{r})]$. The decoded (predicted) orientation was considered correct if it matched the true orientation.

**Supplemental References**